\documentclass[aps,prd,twocolumn,showpacs,amsmath,nofootinbib]{revtex4-1}

\usepackage{color}
\newcommand{\beqa}{\begin{eqnarray}}
\newcommand{\eeqa}{\end{eqnarray}}
\usepackage{float}
\usepackage{graphicx}
\usepackage{latexsym}
\usepackage{amsmath}
\usepackage{hyperref}
\usepackage{mathrsfs}
\usepackage{dcolumn}
\usepackage{bm}%
\usepackage{grffile}
\usepackage{upgreek}
\usepackage{siunitx}

\begin{document}

\title{Constraining spinning primordial black holes with interstellar dust heating}

\author{Qianyong Li}
\author{Yupeng Yang}\email{ypyang@aliyun.com}
\affiliation{ School of Physics and Physical Engineering, Qufu Normal University, Qufu, Shandong, 273165, China }

\begin{abstract}

Primordial black holes (PBHs) are a well-motivated dark matter candidate, and their cosmic abundance is constrained by a variety of observational probes. PBHs in the mass range $10^{15}\,\text{g}\,{-}\,10^{17}\,\text{g}$ are evaporating today via Hawking radiation, a process that can heat interstellar dust and modify its thermal emission. Recent studies have used 
this effect to place constraints on the abundance of non-spinning PBHs. 
We extend this approach by investigating the influence of PBH spin on dust-heating constraints. Furthermore, we account for secondary photons that originate not only from the decay of gauge bosons but also from the decay of hadrons produced via the fragmentation of primary quarks and gluons emitted through Hawking radiation. By comparing the dust heating rate induced by spinning PBHs with the maximum cooling rate of dust, considering both silicate and graphite grains, we derive new upper limits on the fraction of dark matter 
in the form of PBHs, $f_{\rm PBH}$. Our results show that the constraints depend on both PBH mass and spin. Smaller PBHs with higher spin yield stronger limits. For example, in the cases we investigated, the strongest constraint is $f_{\rm PBH} \sim 1.5 \times 10^{-4}$ for $M_{\rm PBH} = 10^{15}{\rm g}$ and spin parameter $a_{*} = 0.9999$. Although these limits are less stringent than existing constraints in the same mass range, they provide a distinct and complementary approach to constraining the abundance of PBHs.

\end{abstract}


\maketitle

\section{Introduction} 

The existence of dark matter is strongly supported by extensive astronomical evidence; however, its microscopic nature remains a fundamental mystery in modern physics\cite{1992ApJ...392..442G, Tisserand_2007, 2007ApJ...657..810D}. Leading candidates for dark matter include weakly interacting massive particles (WIMPs, such as neutralinos in supersymmetric theories), axions (originally proposed to solve the strong CP problem), and primordial black holes (PBHs)\cite{Monroy_Rodr_guez_2014, Arbey_2020, Aleksandrov_2021, Zhang_2022, 2022arXiv221000072R}. Among these, PBHs have drawn significant attention because they do not require new physics beyond the Standard Model\cite{Raidal_2017, PhysRevD.101.043015, Carr_2025}. This interest has been further strengthened by recent gravitational wave observations of black hole mergers, some of which may originate from merging PBHs~\cite{PhysRevLett.116.201301,PhysRevLett.120.191102}.

PBHs are hypothesized to form in the early universe through the gravitational collapse of primordial density fluctuations~\cite{Sasaki_2018,jusufi2025regularblackholesgravitational}. During the radiation-dominated era, collapse occurs if the density contrast $\delta\rho/\rho$ surpasses a critical threshold, typically $\delta_c \sim 0.3$, overcoming radiation pressure~\cite{Carr_2022,PhysRevLett.125.101101,Laha_2020}. The mass of a PBH formed at time $t$ after the Big Bang scales as $M_{\rm PBH} \sim 10^{15}{\rm g}(t/10^{-23}{\rm s})$~\cite{Carr_2010}. Through Hawking radiation, a PBH has an expected lifetime of $\tau_{\rm PBH} \sim 10^{64} (M_{\rm PBH}/M_\odot)^3 {\rm yr}$. Given the current age of the Universe, $\sim 13.7$ billion years~\cite{Planck:2015fie}, PBHs with masses $M_{\rm PBH} \lesssim 10^{15}{\rm g}$ should have already evaporated, as their lifetimes are shorter than the cosmic age~\cite{Bassett_2001,Choudhary_2021,Cotner_2017}. In contrast, PBHs with $M_{\rm PBH} > 10^{15}{\rm g}$ are still evaporating, continuously injecting high-energy photons and particles into the Universe
\footnote{Note that a black hole’s temperature scales as $T_{\rm BH} \sim (M_{\rm PBH}/10^{10}{\rm g})^{-1}$TeV~\cite{Carr_2010}; thus, PBHs with $M_{\rm PBH} \gtrsim 10^{18}{\rm g}$ have temperatures too low to produce appreciable Hawking radiation. Consequently, heavier PBHs have negligible radiative impact on the Universe.}, and are considered dark matter candidates. 

The abundance of primordial black holes (PBHs), considered dark matter candidate, has been constrained by various astronomical observations~\cite{Carr_2025,Yang:2022nlt,Yang:2023qnl,Yang:2025mkp,Hao:2024hzu,Cole:2024wte,Saha:2024ies,Yang:2024snb,Escriva:2022duf,Halder:2022ijw,MaximKhlopov2010,Yang:2020egn,Yang:2020zcu,Yang:2019bkk,Natwariya:2021xki,Zhao:2025ddy,Xie:2024eug,Yang:2024pfb,Zhang:2023rnp,PhysRevD.106.023030,CDEX:2024xqm,Huang:2024xap,Balaji_2025,Khan_2025,Mittal_2022,Ray_2021}. PBHs with masses $10^{15} \text{g}<M_{\text{PBH}} < 10^{18}\,\text{g}$ can emit particles such as photons and electrons via Hawking radiation. Observations of the extragalactic gamma-ray background (e.g., from the Fermi satellite) and of cosmic-ray electrons (e.g., by AMS-02 and Voyager) can therefore be used to constrain such light PBHs~\cite{Carr_2010,Boudaud:2018hqb,Su:2024hrp}. The energy injected by these emitted particles into the Universe can also influence cosmic evolution, allowing cosmic microwave background (CMB) observations to place further limits on their abundance~\cite{Clark:2016nst}. For heavier PBHs with $M_{\text{PBH}} > 10^{18}\,\text{g}$, the Hawking temperature is too low for significant radiative emission. Their influence on the Universe therefore stems mainly from gravitational effects, such as microlensing and gravitational wave signatures~\cite{Mroz:2024wia,Kalogera:2021bya,2025PDU....5002072H,2025arXiv250905618W,2025PDU....4901991D,2025arXiv250707665Y}. In the case of even more massive PBHs, e.g., those exceeding $10\,M_{\odot}$, accretion of baryonic matter can produce radiation analogous to that of astrophysical black holes, but occurring much earlier in cosmic history. This accretion radiation can affect observables including the CMB and the global 21cm signal~\cite{Serpico:2020ehh,Chatterjee:2026nmb,Yang:2021idt}, thereby providing additional constraints on the abundance of PBHs.  
Note that although primordial black holes (PBHs) with masses \( M_{\text{PBH}} < 10^{15}\text{g} \) have evaporated by the present epoch and thus cannot constitute dark matter\footnote{Recently, the ``memory burden'' effect has been proposed~\cite{Dvali:2020wft,Dvali:2018xpy,Dvali:2024hsb,Thoss:2024hsr}, which slows Hawking radiation after a light PBH loses, e.g., half of its mass. This prolongs the lifetime of such PBHs, potentially allowing them to survive until the present and serve as dark matter candidates.}, their earlier high temperature and strong Hawking radiation had significant impacts on cosmic evolution. For example, Big Bang nucleosynthesis provides robust constraints on these light PBHs~\cite{Carr_2025,Carr_2010,Boccia:2024nly}.

Interstellar dust constitutes about 1\% of the interstellar gas mass and is composed predominantly of silicate and graphite grains, with a size distribution often described by the Mathis-Rumpel-Nordsieck (MRN) power law~\cite{1977ApJ...217..425M}. Dust temperatures vary widely with environment: 10-20 K in atomic and molecular regions, 30-200 K in HII zones, and up to 1000-1500 K in circumstellar envelopes~\cite{Melikhov_2022,Melikhov_2023}. Grains cool via infrared emission, with the cooling rate determined by the absorption efficiency and grain temperature. The radiative cooling of a dust grain can be modeled by integrating the product of its absorption efficiency and the Planck function over all wavelengths. For small grains, stochastic heating by individual photons induces large temperature fluctuations that significantly modify the emission spectrum~\cite{1986ApJ...302..363D,Draine:2000dj,2022FrASS...983288H}.

Several dust models have been developed to reproduce the observed infrared emission. One class of models treats dust as a mixture of polycyclic aromatic hydrocarbons (PAHs), amorphous carbon of various sizes, and amorphous silicates, with size distributions adjusted to match both extinction and emission~\cite{Compi_gne_2010}. Another widely used approach combines amorphous silicates, graphitic grains, and PAHs, where the small PAH population is essential for reproducing the well-known aromatic infrared bands 
in the 3-12 $\mu \rm m$ range~\cite{2007ApJ...657..810D}. These models successfully explain dust emission from the diffuse interstellar medium, photodissociation regions, and external galaxies. The composition, size distribution, and temperature of dust grains collectively determine the shape of the infrared spectral energy distribution and thus serve as the basis for inferring the heating sources responsible for the observed emission.

Beyond photon heating, other mechanisms can dominate in specific environments. In hot, tenuous gas with a weak radiation field, electron collisional heating can surpass photon absorption and significantly affect the dust spectral energy distribution~\cite{2013A&A...556A...6B}. Cosmic rays also contribute to dust heating, especially in dense molecular clouds where UV radiation is heavily attenuated; they not only deposit energy into grains but also alter grain charging through secondary electron emission, influencing coagulation and survival~\cite{Ivlev:2015zsa,Kalv_ns_2022}.

Recently, the authors of~\cite{Melikhov_2022} proposed a novel constraint on primordial black holes (PBHs) based on the thermal interaction between the diffuse radiation field from PBH evaporation and interstellar dust. The key point is that if PBHs were overly abundant, the resulting dust heating rate would exceed the infrared cooling rate of dust, leading to anomalously high dust temperatures. This outcome would contradict the widely observed cold dust component and thereby set an upper limit on the PBH abundance. 
In this work, we revisit and extend the previous analysis. In particular, we incorporate the energy spectrum of Hawking radiation from spinning PBHs, for which the photon emission rate per unit energy and time is enhanced compared to the non-spinning case. Furthermore, our treatment of Hawking radiation includes secondary photons, which originate not only from the decay of gauge bosons but also from hadron decays following the fragmentation of primary quarks and gluons emitted via Hawking radiation. These important aspects have not been systematically accounted for in earlier studies.

This paper is structured as follows. In Section II, we introduce the fundamental concepts of PBHs and interstellar dust, outline the methodology employed in this study, and present constraints on the fraction of dark matter that could exist in the form of PBHs. Our conclusions are summarized in Section III.

\section{The basic properties of primordial black holes, the interstellar dust, and final constraints}
\label{sec:basic}

\subsection{The basic properties of primordial black holes}

According to black hole theory, primordial black holes (PBHs) are fundamentally similar to astrophysical black holes. Consequently, established theoretical frameworks for astrophysical black holes can be directly applied to PBHs. In this section, we briefly review the basic properties of PBHs, and further details can be found in, e.g., Refs.~\cite{Carr_2025,Carr_2010,Carr:2009jm}. 

For a non-spinning Schwarzschild primordial black hole~\footnote{In this work, we consider uncharged PBHs.}, the Hawking temperature ($T_{\rm PBH}$) is inversely proportional to its mass ($M_{\rm PBH}$) and is given by \cite{Hawking:1974rv,1984ucp..book.....W,1974Natur.248...30H,Hawking:1975vcx,Carr_2010}

\begin{equation}
T_{\rm PBH} = \frac{\hbar c^3}{8\pi k_{B}GM_{\rm PBH}},
\end{equation}
where $\hbar$, $c$, $G$, and $k_B$ denote the reduced Planck constant, speed of light, gravitational constant, and Boltzmann constant. Previous studies have typically focused on non-spinning PBHs, since theoretical arguments suggest that PBHs generally form with negligible or low spin~\cite{Chiba:2017rvs,Mirbabayi:2019uph,DeLuca:2019buf}. However, a number of formation scenarios allow for PBHs to be born with high spin~\cite{KHLOPOV1980383,Harada:2017fjm,Cotner:2018vug,Arbey:2019jmj,He:2019cdb,Cotner:2019ykd,Dvali:2021byy}. When rotation is taken into account, the black hole temperature becomes~\cite{PhysRevD.13.198,PhysRevD.14.3260,PhysRevD.41.3052,MacGibbon:2007yq}

\begin{equation}
T_{\rm PBH,s}=\frac{\hbar c^3}{4\pi k_{B}G M_{\rm PBH}}\frac{\sqrt{1-a^{2}_{*}}}{1+\sqrt{1-a^{2}_{*}}}
\end{equation}
with the reduced spin parameter $a_{*}=J/(GM^{2}_{\rm PBH})$ and $J$ being the angular momentum of the PBH. The photon emission rate (per unit energy and per unit time) for a 
spinning PBH is written as~\cite{Arbey:2019vqx,PhysRevD.13.198,PhysRevD.14.3260}

\begin{equation}
\frac{\mathrm{d}N_{\gamma}}{\mathrm{d}t \, \mathrm{d}E} {\bigg|}_{{\rm spin}=a_{*}}= \frac{\Gamma (a_{*})}{2\pi\hbar} \left[ \exp\left(\frac{E}{k_B T_{\rm PBH,s}(a_{*})}\right) - 1 \right]^{-1},
\end{equation}
where $\Gamma$ is the photon greybody factor. Besides the photons emitted directly from PBHs via Hawking radiation (referred to as `primary' photons), `secondary' photons are also produced. These originate not only from the decay of gauge bosons, but also from the decay of hadrons generated through the fragmentation of primary quarks and gluons emitted in the Hawking process. Therefore, the total photon spectrum from a PBH can be expressed as~\cite{Carr_2010}

\begin{equation}
\frac{\mathrm{d}N_{\gamma}}{\mathrm{d}t \, \mathrm{d}E} {\bigg|}_{\rm total} = \frac{\mathrm{d}N_{\gamma}}{\mathrm{d}t \, \mathrm{d}E} {\bigg|}_{\rm primary} + 
\frac{\mathrm{d}N_{\gamma}}{\mathrm{d}t \, \mathrm{d}E} {\bigg|}_{\rm secondary}.
\end{equation}

In this work, the photon spectra (primary and secondary) are computed with the public code \texttt{BlackHawk\_v2.3}~\footnote{https://blackhawk.hepforge.org/}, configured with \texttt{hadronization\_choice=3}. This choice selects the \texttt{Hazma} Python library as the internal interface for modeling hadronization processes~\cite{Coogan:2019qpu}. Further details can be found in the \texttt{BlackHawk\_v2.3} manual~\cite{Arbey:2019mbc,Arbey:2021mbl}.

The radiation flux from PBHs at redshift $z$ with energy $E$ can be written as

\begin{equation}
F_{\rm PBH}(E,z)=\frac{c}{4\pi}u(E,z),
\end{equation}
where $u$ is the energy density. Photons emitted at redshift $z$ with energy $E_{z}$ are redshifted to an observed energy today ($z=0$) of $E_0 = E_{z}/(1+z)$. Integrating over energy then yields the energy density at $z=0$ as follows:

\begin{align}
&u(E,z=0) = \frac{u(E,z)}{(1+z)^3} \nonumber \\ 
         &= n_{\rm PBH,0} \int dt \int (1+z)^{2} E_{0} \frac{dN_\gamma}{dt\,dE}\bigl(E_0(1+z)\bigr)\, dE,   
\end{align}
where $n_{\rm PBH,0}$ is the PBH number density at $z=0$. The radiation flux from PBHs at $z=0$ can therefore be 
written as~\cite{Carr_2010,Arbey_2020,2020PhLB..80835624B}

\begin{align}
F_{\rm PBH} 
&= \frac{c}{4\pi}\frac{f_{\rm PBH} \Omega_{\rm DM}\rho_{\rm cri}}{M_{\rm PBH}} 
   \int_{z_{\rm rec}}^0 \frac{(1+z)\,dz}{H(z)} \notag \\
&\quad \times \int_0^\infty E_0 \, \frac{dN_\gamma}{dt\,dE}\bigl(E_0(1+z)\bigr)\, dE,
\label{eq:F_heatpbh}
\end{align}
where $\rho_{\rm cri}$ is the critical density of the Universe, and we have used $dt/dz=-1/[(1+z)H(z)]$ with $H(z)=H_{0}\sqrt{\Omega_{m}(1+z)^{3}+\Omega_{\Lambda}+\Omega_{r}(1+z)^{4}}$, 
$H_0$ is the Hubble constant, and $\Omega_{\rm m}$, $\Omega_\Lambda$, and $\Omega_{\rm r}$ are the density parameters for matter, 
dark energy, and radiation, respectively. 

The quantity $f_{\rm PBH}=\Omega_{\rm PBH}/\Omega_{\rm DM}$ denotes the fraction of dark matter made up of PBH. 
The upper limit in redshift is taken to be the cosmic recombination epoch, $z_{\rm rec} \approx 1100$, when the Universe becomes optically thin to high energy photons. This derivation assumes a monochromatic PBH mass distribution, meaning all PBHs share the same mass. However, our results can be straightforwardly generalized to arbitrary PBH mass functions.

\subsection{The basic properties of interstellar dust}

In this section, we briefly review the basic properties of interstellar dust. Further detailed discussion can be found in, e.g., Refs.~\cite{Dwek:2004ad,Tielens_2005,Melikhov_2022,1977ApJ...217..425M}. 

Interstellar dust constitutes about 1\% of the interstellar gas mass. It forms in stellar atmospheres and nebulae, then cools and acquires icy mantles and electric charges upon entering the interstellar medium. Dust absorbs and reddens starlight, while its infrared emission reveals physical conditions and plays an active cooling role in star formation. Photon absorption heats the grains, causing continuous thermal emission that converts stellar ultraviolet into infrared radiation~\cite{Dwek:2004ad,Tielens_2005}. While its exact nature remains debated, a leading theoretical description is the Mathis-Rumpel-Nordsieck (MRN) model~\cite{1977ApJ...217..425M}. 
The MRN model was originally constructed from extinction observations of the diffuse 
interstellar medium in the Milky Way, and is therefore most suitable for this environment. Nevertheless, it has been successfully extended to a broader range of environments, including different regions of the Milky Way as well as the Large and Small Magellanic Clouds~\cite{Weingartner:2001qu,1992ApJ...395..130P,1994A&A...284..241M}.
 This model treats interstellar dust as a mixture of spherical graphite and silicate grains, with a size distribution following a power law $n(a) \propto a^{-3.5}$ over $0.005 < a < 0.25 \mu m$. Following Ref.~\cite{Melikhov_2022}, we adopt a composition dominated by silicates and graphite, with grain radii ranging from $a = 0.01$ to $0.25\mu\mathrm{m}$, noting that smaller grains exhibit sharp temperature fluctuations due to their low heat capacity.

The temperature of interstellar dust is not uniform. It varies widely across galactic regions~\cite{Dwek:2004ad,Tielens_2005}. In atomic and molecular hydrogen regions, it can drop to 10-20 K, and even to 6 K in dense clouds shielded from starlight. Within HII zones, temperatures typically range from 30-200 K, while the highest values, up to 1000-1500 K, occur in circumstellar envelopes. Based on Refs.~\cite{Melikhov_2022,Tielens_2005}, the dust temperature is estimated as $T_{\rm sil}=13.6(a/\rm {\mu m})^{-0.06}\rm K$ for silicate dust, and $T_{\rm gra}=15.8(a/\rm {\mu m})^{-0.06}\rm K$ for graphite dust. Note that even though we have adopted these two dust temperature values for our calculations, uncertainties remain. 
For instance, the authors of~\cite{Horn:2007kz} investigated equilibrium temperature fluctuations of interstellar dust grains 
and found that the fluctuations decrease with increasing grain size. For our adopted lower size cutoff $a = 0.01\,\mu\mathrm{m}$, 
the fluctuations are at most a few K. For grains with radii $a\gtrsim 0.01\,\mu\mathrm{m}$, the equilibrium 
temperature approximation is therefore justified. Stochastic heating effects become significant only for grains smaller than 
$\sim 0.01\,\mu\mathrm{m}$~\cite{Horn:2007kz}. Thus, the impact on our final constraints is negligible, and our adopted treatment 
remains conservative and does not affect our main conclusions.

The radiative cooling rate of a dust grain can be modeled assuming its emission approximates a modified blackbody spectrum. The rate depends on the grain temperature $T_d$, size $a$, and material properties~\cite{Melikhov_2022}:

\begin{equation}
\frac{dE}{dt} \bigg|_{\rm rad, dust} = 4\pi \sigma_d \int_0^\infty Q_\lambda(a) B_\lambda(T_d)\, d\lambda,
\label{eq:cooling_rate}
\end{equation}
where $B_\lambda(T)$ is the Planck function, and $Q_\lambda(a)$ is the absorption efficiency with $Q_{\lambda}(a) =1$ for $\lambda \leq 2\pi a$ and 
$Q_{\lambda}(a) =2\pi a/\lambda$ for $\lambda > 2\pi a$, and $\sigma_{d} = \pi a^{2}$ is the geometric cross section. To derive conservative upper limits, we focus on the smallest grains ($a=0.01~\mu{\rm m}$), which have the lowest heat capacity. Using the temperature estimates $T_{\rm sil}$ and $T_{\rm gra}$ from above, the corresponding maximum radiative cooling rates are $(dE/dt)_{\rm sil}=2.76\times 10^{14}{\rm erg~s^{-1}}$ for silicate and $(dE/dt)_{\rm gra}=5.84\times 10^{14}{\rm erg~s^{-1}}$ for graphite, as given in Ref.~\cite{Melikhov_2022}. Interstellar dust is heated by radiation from various astrophysical sources, including primordial black holes. Assuming the PBH radiation is isotropic and the dust grain absorbs all incident radiation within its geometric cross section, the heating rate due to PBHs is given by~\cite{Melikhov_2022}

\begin{equation}
\frac{dE}{dt} \bigg|_{\rm heat, PBH} = 4\pi \sigma_d F_{\rm PBH},
\end{equation}
where the isotropic PBH radiation flux $F_{\rm PBH}$ is calculated from Eq.~(\ref{eq:F_heatpbh}).

We further emphasize the issue of the dust reaching thermal equilibrium. The validity of the thermal equilibrium assumption depends strongly on grain size. For very small grains with radii smaller than about \(0.01\)-\(0.02\,\mu\mathrm{m}\), the heat capacity is so low that the absorption of a single photon can cause large temperature spikes, and such grains do not possess a well-defined equilibrium temperature 
\citep{Horn:2007kz}. In contrast, for larger grains with radii greater than about \(0.03\)-\(0.05\,\mu\mathrm{m}\), the heat capacity is 
sufficiently large that the temperature change due to a single photon absorption is small, and the thermal equilibrium approximation is 
well justified \citep{1986ApJ...302..363D,Draine:2000dj}. Our adopted lower size cutoff \(a = 0.01\,\mu\mathrm{m}\) lies in the transition region; 
although grains of this size still experience some fluctuations, the amplitude is at most a few K \citep{Horn:2007kz}. Therefore, applying 
the equilibrium temperature approximation to grains with \(a\gtrsim 0.01\,\mu\mathrm{m}\) is reasonable and conservative.


 \subsection{Constraints on the abundance of PBHs}

As discussed in the preceding sections, interstellar dust radiates at a temperature that depends on its size and is also subject to heating by astrophysical sources. To derive conservative upper limits, we consider primordial black holes (PBHs) as the sole heating source. The key criterion for obtaining these limits is that the heating rate from PBH Hawking radiation not exceed the radiative cooling rate of a dust grain. In practice, the constraint on PBHs can be expressed by the condition:

\begin{equation}
\frac{dE}{dt}(f_{\rm PBH}) \bigg|_{\rm heat, PBH} \leq \frac{dE}{dt} \bigg|_{\rm rad, dust}.
\label{eq:condition}
\end{equation}

\begin{figure}
    \centering
    \includegraphics[width=0.55\textwidth]{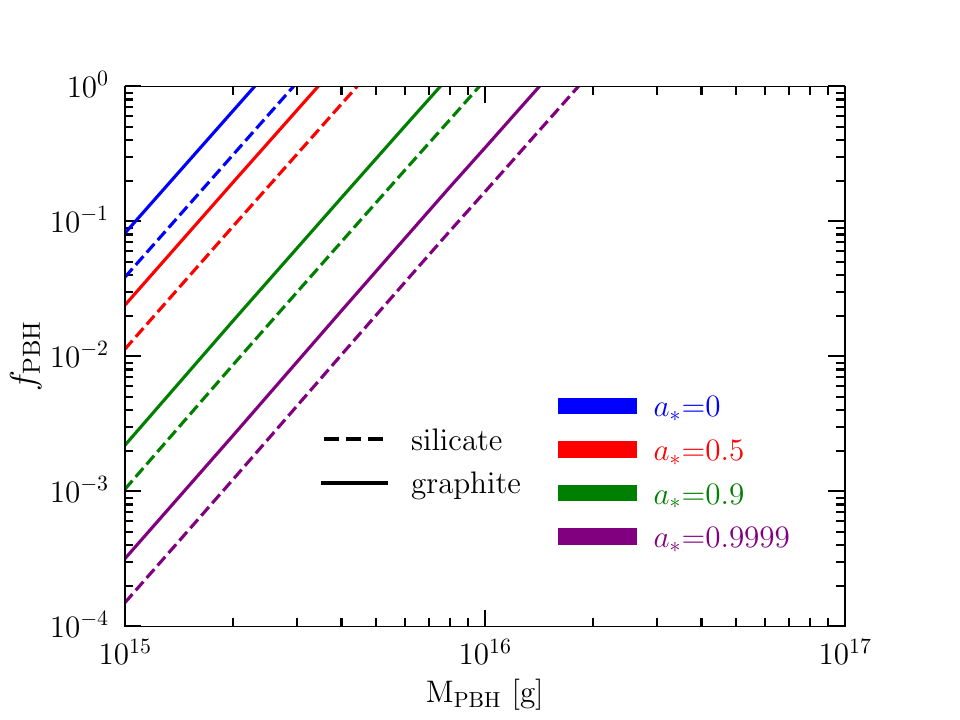}
    \caption{Upper limits on the fraction of dark matter in PBHs, 
    $f_{\rm PBH}=\Omega_{\rm PBH}/\Omega_{\rm DM}$, derived from interstellar dust heating constraints. The curves (from top to bottom) 
    correspond to PBH spin parameters $a_{*}=0,0.5,0.9$, and 0.9999. Dashed and solid lines represent results for silicate and graphite dust grains, respectively.}
    \label{fig:spin_limits}
\end{figure}

The resulting upper limits on the fraction of dark matter in the form of PBHs, $f_{\rm PBH}=\Omega_{\rm PBH}/\Omega_{\rm DM}$, are presented in Fig.~\ref{fig:spin_limits} for both silicate and graphite dust, considering spin parameters $a_{*}=0, 0.5, 0.9$, and 0.9999. For a fixed spin parameter $a_{*}$, the limit is stronger for silicate dust than for graphite, due to its lower equilibrium temperature. For a given dust composition, the constraints become more stringent with higher spin parameters, because higher-spin PBHs emit more photons. Among the cases examined, the strongest upper limit obtained is $f_{\rm PBH} \sim 1.5 \times 10^{-4}$ for $M_{\rm PBH} = 10^{15}{\rm g}$ and spin $a_{*} = 0.9999$. 

We note that our results for non-spinning PBHs ($a_{*}=0$) differ from those in Ref.~\cite{Melikhov_2022} due to our use of a different energy spectrum for Hawking radiation. It should also be noted that in this work we do not account for the change in PBH mass during Hawking evaporation. Over the PBH mass range considered, mass loss due to Hawking radiation has noticeable effects for the lighter PBHs. Consequently, it is expected that the constraints would become stronger at the lower end of the considered PBH mass range (around $10^{15}\rm g$). Furthermore, the limits presented here are conservative, as we have set the dust grain size to $a=0.01\ \mu\mathrm{m}$, corresponding to the lower bound of the size range we consider. For a larger grain size (e.g., $a=0.25\ \mu\mathrm{m}$), the equilibrium dust temperature is lower, leading to a reduced radiative cooling rate. To satisfy Eq.~(\ref{eq:condition}), the allowable heating rate must be correspondingly lower, resulting in stronger constraints on $f_{\rm PBH}$.

Other studies have placed stringent limits on spinning PBHs using different probes. The authors of~\cite{Dasgupta:2019cae} constrained spinning PBHs using observations of the diffuse supernova neutrino background (DSNB) by Super-Kamiokande and the 511 keV gamma-ray line from positron annihilation in the Galactic center as measured by INTEGRAL. They considered spin parameters of $a_{*}=0,0.5,0.9$, and 0.9999. The strongest upper limit on the PBH abundance for a monochromatic mass function is $f_{\rm PBH}\sim 4\times 10^{-4}$ at $M_{\rm PBH}=10^{15}\rm g$ for $a_{*}=0.5$ from the DSNB analysis. For the 511 keV line, the strongest limit is $f_{\rm PBH}\sim 4\times 10^{-4}$ at $M_{\rm PBH}=2 \times 10^{16}\rm g$ for $a_{*}=0.5$. In~\cite{Arbey:2019vqx}, the authors investigated the fraction of dark matter in PBHs using the isotropic gamma-ray background. For PBH masses in the range $10^{15} < M_{\rm PBH} < 10^{17} \rm g$, the strongest upper limit was found to be $f_{\rm PBH}\sim 6\times 10^{-8}$ at $M_{\rm PBH}=10^{15}\rm g$ for $a_{*}=0.9$, with a nearly identical limit for $a_{*}=0.9999$. The authors of~\cite{Natwariya:2021xki} derived constraints on spinning PBHs from the global 21cm differential brightness temperature, requiring that the deviations from the standard cosmology prediction do not exceed a factor 1/4. They obtained upper limit of $f_{\rm PBH}\sim 4\times 10^{-10}$ ($6\times 10^{-10}$) for $a_{*}=0.9999$ (0.9) at $M_{\rm PBH}=10^{15}\rm g$. A common trend across these studies, consistent with our findings, is that higher spins lead to more stringent constraints, narrowing the mass window where PBHs could constitute all of dark matter.

\begin{figure}
    \centering
    \includegraphics[width=0.55\textwidth]{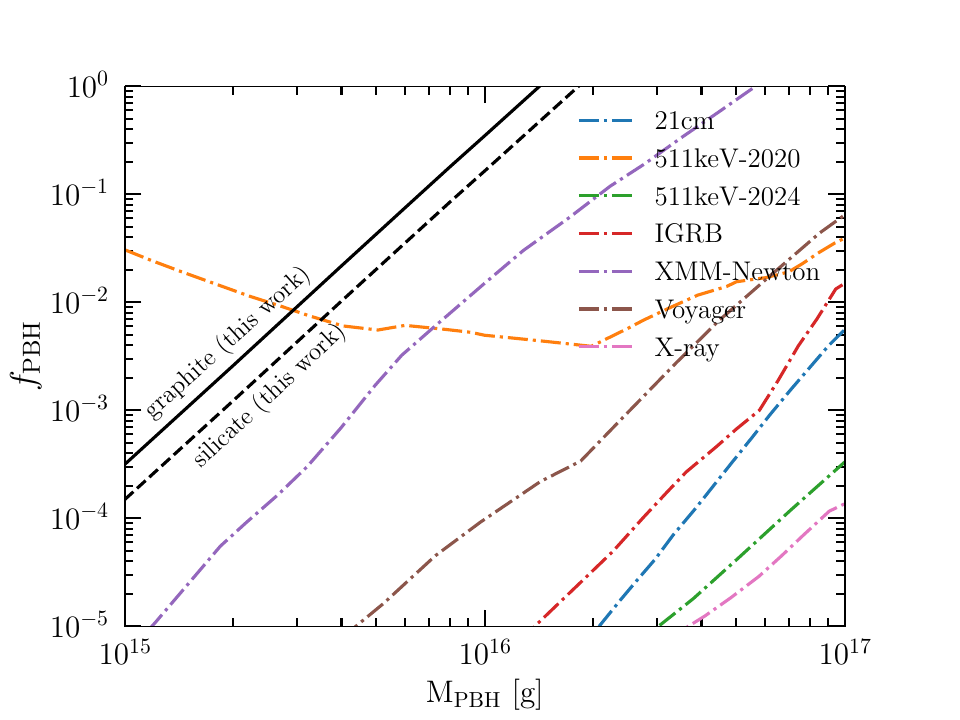}
    \caption{Comparison of constraints on the PBH abundance, $f_{\rm PBH}=\Omega_{\rm PBH}/\Omega_{\rm DM}$, 
    from various astrophysical and cosmological observations. The limits from dust heating for silicate (dashed line) and graphite (solid line) grains are shown in black, for a PBH spin parameter $a_{*}=0.9999$. For comparison. we also include constraints 
    from several other probes with $a_* = 0.9999$ (dot-dashed lines): 
(1) constraints from the global 21 cm differential brightness temperature (21cm)~\cite{Natwariya:2021xki};
(2) constraints from the Galactic center 511 keV gamma-ray line using the conservative method (511keV-2020)~\cite{Laha:2019ssq}; 
(3) constraints from the Galactic center 511 keV gamma-ray line using the comprehensive method (511keV-2024)~\cite{DelaTorreLuque:2024qms}; 
(4) constraints from the isotropic gamma-ray background (IGRB)~\cite{Arbey:2019vqx}; 
(5) constraints from diffuse X-ray observations with XMM-Newton (XMM-Newton)~\cite{DelaTorreLuque:2024qms}; 
(6) constraints from Voyager spacecraft measurements of positrons and electrons (Voyager)~\cite{DelaTorreLuque:2024qms}; 
(7) constraints from the cosmic X-ray background (X-ray)~\cite{Tan:2024nbx}.
}
    \label{fig:comparison}
\end{figure}

For comparison with established constraints, we plot several existing limits for $a_* = 0.9999$ 
in Fig.~\ref{fig:comparison}: 
(1) limits from the global 21 cm differential brightness temperature (21cm)~\cite{Natwariya:2021xki};
(2) conservative limits from the Galactic center 511 keV gamma-ray line (511keV-2020)~\cite{Dasgupta:2019cae}; 
(3) comprehensive limits from the Galactic center 511 keV gamma-ray line (511keV-2024)~\cite{DelaTorreLuque:2024qms}; 
(4) limits from the isotropic gamma-ray background (IGRB)~\cite{Arbey:2019vqx}; 
(5) limits from diffuse X-ray observations with XMM-Newton (XMM-Newton)~\cite{DelaTorreLuque:2024qms}; 
(6) limits from Voyager spacecraft measurements of positrons and electrons (Voyager)~\cite{DelaTorreLuque:2024qms}; 
(7) limits from the cosmic X-ray background (X-ray)~\cite{Tan:2024nbx}. 
In this figure, we present our strongest limits (for $a_{*}=0.9999$) for both silicate and graphite dust. 
Our silicate constraints at $a_* = 0.9999$ are stronger than those from the conservative limits from the 
Galactic Center 511 keV gamma-ray line~\cite{Dasgupta:2019cae} for PBH masses $M_{\rm PBH}<3\times10^{15}\rm g$, but weaker than those from 
the comprehensive method~\cite{DelaTorreLuque:2024qms}. Overall, our constraints are weaker than the other existing limits. 
Nonetheless, the dust heating method provides a distinct and complementary approach to constraining PBH abundance.

We note that in this work we consider PBHs as an additional heating source for interstellar dust, on top of the dominant heating from astrophysical sources such as stars. Since the inclusion of stellar heating would further raise the dust temperature, the allowable PBH fraction required to reach the same cooling threshold would be lower. Therefore, the constraints on $f_{\mathrm{PBH}}$ derived in this work should be regarded as conservative upper limits. We also note that the heating flux used here is computed from cosmologically distributed PBHs. In principle, PBHs distributed in the dark matter halo of the Milky Way could also contribute to the local heating flux.\footnote{We have explored this scenario in a companion paper, and find that the resulting constraints on $f_{\mathrm{PBH}}$ are moderately stronger than those presented here.} Including both cosmological and halo contributions would further strengthen the limits. Finally, we emphasize that our current constraints are based on dust in the Milky Way. Targeting environments with a lower interstellar radiation field, where the dust cooling rate is correspondingly reduced, would yield stronger constraints on $f_{\mathrm{PBH}}$. We plan to investigate such targets, e.g., high-latitude diffuse clouds or dwarf galaxies, in future work.

In this work, we adopt a uniform distribution for interstellar dust throughout the Galaxy, a common assumption in 
studies of diffuse interstellar dust emission where the dust properties are characterized by observationally determined 
average temperatures. The PBH heating flux is taken as the cosmological background, which is isotropic and homogeneous on large scales. 
Since the heating source itself is isotropic, the potential anisotropic distribution of dust due to a spatially varying PBH density 
would have a negligible impact on the average heating rate per unit dust mass when averaged over the Galactic disk. 
However, if one considers the additional contribution from PBHs distributed within the Milky Way dark matter halo, 
the non-uniform dust distribution would become more relevant, as regions of higher PBH density 
(e.g., the Galactic center) would experience enhanced local heating 
\footnote{This scenario has been discussed in a separate work, and we plan to investigate it in more detail in future studies.}. 
For the present analysis, which focuses on the cosmological PBH background as the dominant heating source, 
the assumption of a uniform dust distribution is well justified and yields conservative limits.

Note that in addition to photons, PBH Hawking radiation also emits electrons and positrons. Since we consider the cosmological contribution 
of PBHs integrated over redshifts from \(z\sim 0\) to \(z\sim 1100\), these charged particles cannot travel such vast distances to the 
present-day Galaxy, unlike photons, due to their strong interactions with the intergalactic and interstellar media. Furthermore, although 
inverse Compton scattering of cosmic microwave background photons could in principle occur, the energies of electrons and positrons 
emitted by PBHs in the mass range considered here are too low to efficiently upscatter CMB photons to energies that would contribute 
to dust heating. We therefore neglect the heating contribution from electrons and positrons in this work.

\subsection{Comparison with other dust models}

Our baseline dust model adopts the Mathis - Rumpel - Nordsieck (MRN) size distribution, a simple power-law $dn/da \propto a^{-3.5}$ for graphite and silicate grains over $0.005<a<0.25\,\mu\mathrm{m}$~\cite{1977ApJ...217..425M}. However, other widely used models include a significantly larger population of very small carbonaceous grains (PAHs) than the MRN model~\cite{Weingartner:2001qu,2007ApJ...657..810D,Compi_gne_2010,Hensley_2023}. It is important to assess how this affects our PBH constraints.

For PBHs with masses $10^{15}<M_{\rm PBH}<10^{17}\,\mathrm{g}$, the Hawking radiation consists of photons with energies $\sim 0.1$-$100\,\mathrm{MeV}$. In this energy range, the dust absorption cross section per unit mass scales as $\propto a^{-1}$, so small grains are much more efficient absorbers than large grains. Models with more small grains therefore have a larger total absorption cross section per unit dust mass, leading to a higher PBH heating rate for a given $f_{\mathrm{PBH}}$.

To quantify this, consider the size distributions constructed to reproduce the observed variations in the interstellar extinction curve~\cite{Weingartner:2001qu}. For the diffuse ISM ($R_V=3.1$), these models contain a carbon abundance of $b_C\sim6\times10^{-5}$ in very small grains. The mass in grains with $a<0.01\,\mu\mathrm{m}$ is about a factor of 3 higher than in the MRN model. Since small grains ($a\sim0.005\,\mu\mathrm{m}$) have a factor of $\sim20$ larger absorption cross section per unit mass than large grains ($a\sim0.1\,\mu\mathrm{m}$), this translates to an overall increase in the dust heating rate by roughly a factor of 2. This would strengthen the PBH constraints (lower $f_{\mathrm{PBH}}$) by a similar factor. A comparable enhancement is expected for models with PAH mass fractions of $4.6\%$ or higher, as constrained by infrared observations of the diffuse interstellar medium~\cite{2007ApJ...657..810D,Compi_gne_2010}. For example, the model of Ref.~\cite{Compi_gne_2010} contains a PAH mass abundance of $7.8\times10^{-4}$ relative to hydrogen, corresponding to an even larger fraction of mass in ultrasmall grains, which would further strengthen the constraints.

The cooling rate is also model-dependent. For large grains, cooling scales with grain volume; for small grains, stochastic heating effects reduce the time-averaged emission efficiency~\cite{Draine:2000dj}. Models with more small grains therefore have a slightly lower cooling rate per unit mass, which partially offsets the increased heating. Including this effect, we estimate that the net shift in the $f_{\mathrm{PBH}}$ upper limits is about a factor of 2 - 3 between the MRN model and the more realistic models.

We therefore conclude that the choice of dust model introduces a systematic uncertainty of roughly a factor of 2 - 3  in our PBH constraints. Our MRN-based results thus provide a conservative benchmark, and future work should explore this model dependence using a grid of dust models to obtain more robust constraints.

\section{Conclusion}
\label{conclusion}

    We have investigated constraints on the fraction of dark matter in primordial black holes (PBHs), $f_{\rm PBH}=\Omega_{\rm PBH}/\Omega_{\rm DM}$, by considering their heating effect on interstellar dust. The dust grains (assumed to be silicate or graphite with radii 
of $0.01$-$0.25~\mu\mathrm{m}$) radiate thermally, approximately as blackbodies, with a cooling rate that depends on their composition, temperature, and size. By requiring that the heating rate from PBH Hawking radiation does not exceed this radiative cooling rate, we derive conservative upper limits on $f_{\rm PBH}$. In contrast to previous analyses, we consistently account for spinning PBHs, with spin parameters $a_{*}=0,0.5,0.9,0.9999$, and include both primary and secondary photon spectra from Hawking radiation. 

We have found that, within the parameter space considered, the strongest limit is $f_{\rm PBH} \sim 1.5 \times 10^{-4}$ for $M_{\rm PBH} = 10^{15}{\rm g}$ and spin parameter $a_{*} = 0.9999$, obtained for silicate dust of size $a=0.01~\mathrm {\mu m}$. The constraints on $f_{\rm PBH}$ become stronger for larger PBH spins, and silicate dust yields tighter limits due to its lower temperature. Although our limits are weaker than some existing constraints from other observations, interstellar dust heating provides a novel and complementary approach for constraining PBHs.

\section*{Acknowledgements}
Y. Yang thanks Junsong Cang for guidance on using the public code \texttt{BlackHawk}. We thank Qiang Yuan for very useful comments. 
We also thank Shuangxi Yi and Yankun Qu for helpful discussions. This work is supported by the Shandong Provincial Natural Science Foundation (Grant No. ZR2025MS16).

\newpage 

\bibliographystyle{apsrev4-1}
\bibliography{ref}

\end{document}